# Imbalanced Classification In Faulty Turbine Data: New Proximal Policy Optimization

Mohammad Hossein Modirrousta, Mahdi Aliyari Shoorehdeli, *Senior Member*, *IEEE*, Mostafa Yari and Arash Ghahremani

***Abstract*—There is growing importance to detecting faults and implementing the best methods in industrial and real-world systems. We are searching for the most trustworthy and practical data-based fault detection methods proposed by artificial intelligence applications. In this paper, we propose a framework for fault detection based on reinforcement learning and a policy known as proximal policy optimization. As a result of the lack of fault data, one of the significant problems with the traditional policy is its weakness in detecting fault classes, which was addressed by changing the cost function. Using modified Proximal Policy Optimization, we can increase performance, overcome data imbalance, and better predict future faults. When our modified policy is implemented, all evaluation metrics will increase by $3\%$ to $4\%$ as compared to the traditional policy in the first benchmark, between $20\%$ and $55\%$ in the second benchmark, and between $6\%$ and $14\%$ in the third benchmark, as well as an improvement in performance and prediction speed compared to previous methods.**

## I. Introduction

PRODUCTION safety and product quality are key factors to ensure economic benefits in the industrial process [1], [2]. Monitoring and diagnosing faults play an important role and have received considerable attention from academia and industry [3], [4].

Operational failures are mainly traced to external environmental factors during the in-service period. Structures can fail due to ice and sand accumulations, erosion, corrosion, lightning damage, dust, and insect contamination [5]. Cracks and delaminations can also occur during the operational phase. When the incipient damage is not detected early, it may result in severe malfunctions, fatal injuries, or, worst case, the collapse of the entire turbine.

Condition Monitoring (CM) systems for turbines rely on fault detection and prediction algorithms, but there are two major challenges. Sensor data generation is the first of these. CM systems and their interpretation are complicated by the need to store and process these data. Sensor reliability is the second issue. [6], [7] have pointed out issues regarding the accuracy and accountability of sensors used for pattern recognition and fault identification. The accuracy of sensor data strongly influences a CM system's performance. Additionally, the use of a large number of sensors, and hence monitoring variables, may reduce the overall reliability of the sensor system. Consequently, fault detection methods are a necessary component of ensuring the operational safety and reliability of mechanical systems and reducing maintenance costs.

Mohammad Hossein Modirrousta (e-mail: mohammadbc@email.kntu.ac.ir) is with the Faculty of Electrical Engineering, K. N. Toosi University of Technology, Tehran, Iran.
Mahdi Aliyari Shoorehdeli (e-mail: aliyari@kntu.ac.ir) is with the Faculty of Mechatronics Engineering, K. N. Toosi University of Technology, Tehran, Iran.
Mostafa Yari (e-mail: yari.mostafa@mapnaec.com) is with the Faculty of Mechatronics Engineering, K. N. Toosi University of Technology, Tehran, Iran.
Arash Ghahremani (e-mail: ghahremani.arash@mapnaec.com) is with the Faculty of Mechanical Engineering, K. N. Toosi University of Technology, Tehran, Iran.

### A. Related work

According to [8], there are three types of fault detection methods for turbine systems, including signal-based methods [9], [10], model-based methods [11], and knowledge-based methods [12], [13]. A robust fault estimation and fault-tolerant control approach for Takagi-Sugeno fuzzy systems is proposed in [14]. Furthermore, multivariate statistical methods have also successfully detected faults [15], [16].

The use of SCADA data for monitoring wind turbine conditions has been proposed in several ways. There has been extensive use of artificial neural networks (ANNs), Gaussian processes (GP), support vector machines (SVMs), and random forests (RFs) in order to improve the performance of turbines. In [17], [18], [19], [20], [21], general overviews of these techniques are provided.

The development of turbine models and fault detection methods based on artificial intelligence has been impressive [22], [23]. As part of the analysis for turbine condition monitoring [24], a method based on artificial neural networks (ANNs) has been proposed. [25] presents an algorithm for ANN pattern recognition and its application to controls for turbines. In [26], support vector machines (SVMs) were combined with a residual-based method to detect and isolate faults in wind turbines. Using Shannon wavelet SVMs and manifold learning to diagnose faults in turbine transmission systems were proposed in [27]. Detecting turbine faults can also be accomplished using AI-based methods based on fuzzy logic or expert systems. According to [28], adaptive neuro-fuzzy inference systems can be used to monitor the condition of turbines.

Deep learning (DL) has achieved enormous success in many fields in recent years. Researchers are also interested in the field of fault detection [29], [30]. An unsupervised deep learning method (denoising autoencoder) has been proposed for detecting turbine faults [31]. In addition to the domain adaptation using the maximum mean discrepancy, DL models such as sparse autoencoder [32] and convolutional neural networks (CNN) [33] are used for condition recognition. An ImageNet pre-trained network is used in the paper [34] to train a deep-learning network to classify faults. In order to obtain a time-frequency distribution to fine-tune the high-level network layers, sensor data are transformed into image data by plotting or using wavelet transformation [35].

An imbalance of class samples occurs in real-world applications when one class's samples exceed those of other classes [36]. Turbine fault detection, for instance, exhibits class imbalance because these machines generally operate under normal conditions and occasionally fail; these conditions result in many normal operations and few faulty ones. Data and algorithm-level research has been conducted to alleviate the problem of class imbalance [37]. This technique is a data-level method that involves random under- and oversampling. Oversampling techniques have been studied extensively [38]. A focus on cost-sensitive algorithms [39] and ensemble learning [40] is discussed. Misclassified positive and negative samples incur high and low costs, respectively. Costs are difficult to determine. An ensemble learning algorithm such as Adaboost [14] uses an iterative boosting algorithm to increase the weight of misclassified samples and decrease the weight of correctly classified samples after each iteration. The performance of boosting depends strongly on the base classifier.

Artificial intelligence (AI) also includes reinforcement learning (RL). Intelligent computing techniques are used to automate and understand problem-oriented learning and decision-making [41]. In contrast to other intelligent methods, it emphasizes that the agent learns through direct interaction with the environment without requiring imitation of supervision signals. Deep reinforcement learning (DRL) combines RL with deep learning (DL). DRL has achieved great success in games [42], recommendation systems [43], and robotics control [44]. Nevertheless, DRL is rarely mentioned in fault identification, which DL dominates. We propose a new method for identifying faults in real turbines through DRL. Intelligent methods will be more universal with fault parameterization and DRL implementation. Several faults are parameterized here, which enables DRL to transform a classification problem into a sequential decision problem.

A Markov decision process for classification was proposed by Wiering et al. This framework defined a standard classification problem as a sequential decision-making problem, and an MLP model trained in it outperformed a regular MLP model trained by backpropagation [45]. Using DRL to identify bearing health states, Ding et al. proposed an approach [46]; Huang et al. adopted DRL to implement a preventive maintenance policy for serial production lines [47]. Based on time-frequency representations (TFR) and dynamic response mappings (DRLs), Wang et al. developed a new fault diagnosis methodology [48]. As described in [49], DDQN improves the detection performance of cyberattacks by adopting a fine-grained traffic flow monitoring mechanism.

In contrast to traditional policy gradient algorithms, PPO is an advanced algorithm capable of overcoming the problem of low learning efficiency caused by the influence of the step size on learning efficiency. PPO has several primary advantages for training control policies. First of all, in [50], the hyperparameters of PPO were proved to be robust when training various tasks, and PPO can balance the complexity and accuracy of control policies. Second, in [51], the training control policy of PPO was found to be superior to that of other RL algorithms on all metrics compared to the performance indicators. Based on the full six-degree-of-freedom system dynamics of the UAV, in [52], PPO is used to train quadrotor control policies, achieving stable hovering. In the context of Software-Defined Networking (SDN), Zolotukhin et al. proposed an interesting approach [53]. The authors investigate Deep Q-Network (DQN) and PPO in response to an attack. DQN and PPO show promising results, which further motivates this study. PPO also simplifies implementation and improves performance in IoT applications [54]. The authors of [55] propose an agent-based reinforcement learning scheme using PPO to allow multiple agents to control their own devices. An intrusion detection hyperparameter control procedure is built in [56], which controls and trains a deep neural network feature extractor based on proximal policy optimization (PPO). A new automated lane change strategy using proximal policy optimization is proposed [57], which shows excellent benefits while maintaining performance stability.

### B. Our Contributions

The findings of the above studies led us to propose a new fault detection method based on DRL. It is based on Classification Markov Decision Process (CMDP), which defines the fault detection problem as a guessing game. The diagnosis agent first learns an optimal recognition policy within the framework of DRL. By using experience replay (ER), the agent automatically interacts with the environment, creating experiences, and updating the model based on those experiences. Furthermore, the proposed method has been tested on detection tasks and is compared to existing detection methods. In addition to exhibiting better generalization and speed compiling, this method also performs well when dealing with imbalanced problems.

The following are the main contributions of our framework as described in this paper:

1) We are using reinforcement learning to build a recommender label system. In order to discover a new algorithm for fault detection from a DRL perspective, we consider fault detection as a guessing game and describe it as a sequential decision-making problem based on CMDP.
2) With the help of experience replay (ER) and reward-based learning tools, an optimal model for fault detection can be developed based on Proximal Policy Optimization (PPO).
3) It is necessary to make some changes to the regular cost function of PPO. As a result of these changes, imbalances in data will be addressed. Using our approach, we can make decisions without relying on feature engineering.
4) We examined and compared the performance of this method with and without changes in the cost function, as

well as with other methods on multiple datasets. We found that this method enhanced fault detection abilities.

## II. Problem definition

The diagnosis of faults is often considered to be a classification problem. The primary purpose of DRL is to solve the sequential decision-making problem. A guessing game is used in this work to diagnose turbine faults. We also create a game simulation that converts fault diagnosis into a sequential decision-making problem using the DRL. This illustration illustrates how a training dataset that might be regarded as a guessing question set is $X_{\text{train}} = \{(s_1, l_1), (s_2 l_2), \ldots, (s_n, l_{\text{n}})\}$ where $s_i$ is the $i - th$ sample and $l_i$ is the $i - th$ label corresponding to sample $s_i$. As part of this game, each round consists of $T$ questions matching training data generated from the training dataset $X_{\text{train}}$ . Agents guess these questions sequentially by the order in which the samples in $D$ are arranged.

This game involves the agent observing a sample each time and assuming the class of the sample. Following the guessing question, the environment provides an immediate reward to the agent and the next guessing question (i.e., the following sample). A positive reward is awarded to the agent if the agent correctly identifies the sample's category; otherwise, a negative reward is given. The agent's objective in this game is to maximize accumulated rewards within the constraints of an optimal behavior policy that has been learned due to constant interaction with the environment, as shown in Fig. 1.

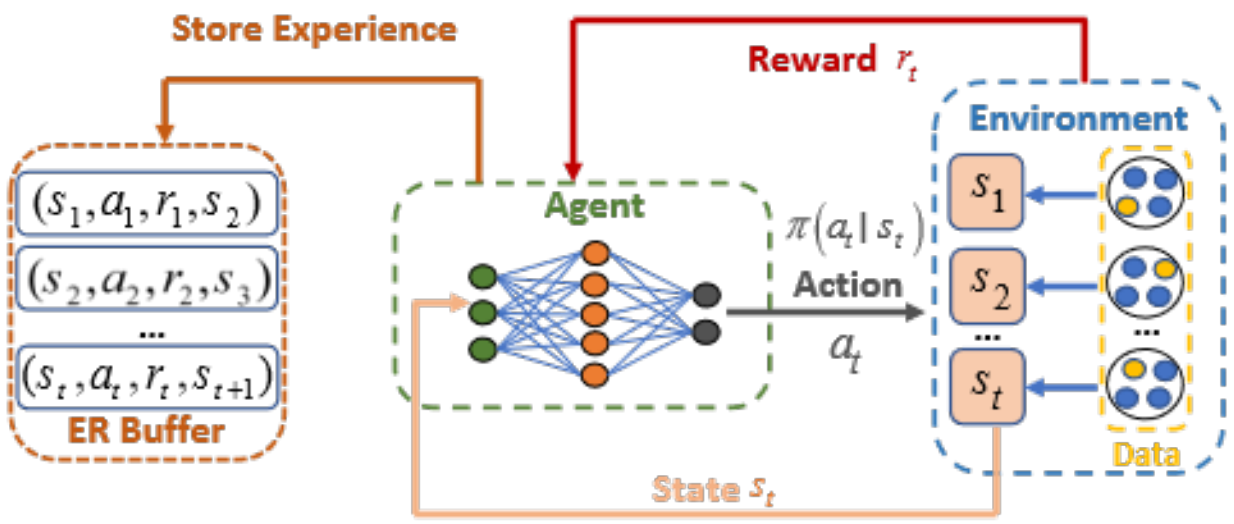

Fig. 1: Overview of the Agent-User interaction in the Classification MDP (CMDP).

## III. Methodology

### A. Proximal Policy Optimization

Using reinforcement learning, an agent can learn how to interact with its environment to maximize its expected cumulative rewards. An RL algorithm can be divided into two general categories: value-based and policy-based. Even though value-based methods can approximate the value function using neural networks in an off-policy manner, policy-based methods, such as the REINFORCE algorithm [20], offer the primary advantage of optimizing the quantity of advantage directly while maintaining stability during the approximation of functions. As a result, our study focuses on RL methods based on policy.

During a general policy gradient reinforcement learning, the objective function is as follows:

$$L^P(\theta) = \hat{\mathbb{E}}_t \left[ \log \pi_\theta (a_t \mid s_t) \hat{A}_t \right] \tag{1}$$

In this case, $\hat{\mathbb{E}}_t$ represents the expectation operator, $\pi_\theta$ represents a stochastic RL policy, and $\hat{A}_t$ represents the estimated advantage function at time step $t$.

Using the discount factor $\gamma \in [0, 1]$, we can calculate $\hat{A}_t$ using the generalized advantage estimator [30]. Generally, the generalized advantage estimator can be described as follows:

$$\hat{A}_t = \delta_t + (\gamma\lambda)\delta_{t+1} + \cdots + \cdots + (\gamma\lambda)^{T-t+1}\delta_{T-1} \tag{2}$$

Where $\delta_t = r_t + \gamma V_\phi(s_{t+1}) - V_\phi(s_t)$, and $T$ is the sampled mini-batch size. The parameter $\lambda \in [0, 1]$ represents the generalized advantage estimator.

According to $L^V$, the objective function is as follows:

$$L^V(\phi) = \widehat{E}\left[L_t^V(\phi)\right] = \widehat{E}\left[\left|\hat{V}_\phi^{\text{target}}(s_t) - V_\phi(s_t)\right|\right] \tag{3}$$

In this case, the target value of the time-difference error (TD-Error) is

$$\hat{V}_\phi^{\text{target}}(s_t) = r_{t+1} + \gamma V_\phi(s_{t+1}) \tag{4}$$

A gradient descent algorithm is used to update the parameters of $V_\phi$, with the gradient $\nabla L^V$:

$$\phi = \phi - \eta_\phi \nabla L^V(\phi) \tag{5}$$

In the critic model optimization, $\eta_\phi$ is the learning rate.

Using the actor model, the PPO uses the importance sampling method in place of the objective function presented in Equation (1) to estimate the expectation of samples collected from the old policy $\pi_{\theta_{old}}$ under the new policy $\pi_\theta$. Using $L^{CPI}$ as a surrogate objective function, the algorithm maximizes:

$$L^{CPI}(\theta) = \widehat{E}_t \left[ \frac{\pi_\theta(a_t \mid s_t)}{\pi_{\theta_{old}}(a_t \mid s_t)} \widehat{A}_t \right] \tag{6}$$

The PPO optimizes $L^{CPI}$ with a small value $\delta$ based on the amount of the policy update as a constraint, according to the equation below:

$$\widehat{E}_t \left[ KL \left[ \pi_{\theta_{\text{old}}}(\cdot \mid s_t), \pi_\theta(\cdot \mid s_t) \right] \right] \leq \delta \tag{7}$$

Kullback-Leibler divergence is indicated by $KL$ [32]. The actor-critic structure is used in advanced policy-based algorithms, including TRPO [18] and PPO [17], which combine the advantages of traditional value-based and policy-based approaches. As well as being more straightforward to implement and allowing multiple optimization iterations, the PPO algorithm also has a higher sample complexity. In particular, PPO proposes a modified surrogate loss function, which combines the policy surrogate and the error term associated with the value function, defined as follows [17]

$$L_t^{CLIP+V+S}(\theta) = \hat{\mathbb{E}}_t \left[ L_t^{CLIP}(\theta) - c_1 L_t^V(\theta) + c_2 S[\pi_\theta](s_t) \right] \tag{8}$$

In this equation, $L_t^{CLIP}$ stands for the clipped surrogate objective, $c_1$ and $c_2$ represent the coefficients, $L_t^V$ stands for the squared-error loss of the value function, and $S$ stands for the entropy loss. In order to ensure that our agents have enough exploration, we use an entropy term. By using this term, the policy will be pushed to behave more spontaneously until the other objective overtakes it.

In more detail, the clipped surrogate objective $L_t^{CLIP}$ is as follows:

$$L^{CLIP}(\theta) = \widehat{E}_t \left[ L_t^{CLIP}(\phi) \right] = \widehat{E}_t \left[ \min \left( R_t(\theta), \operatorname{clip} \left( R_t(\theta), 1-\epsilon, 1+\epsilon \right) \right) \widehat{A}_t \right] \tag{9}$$

Where $\epsilon$ indicates a clipping parameter and $R_t(\theta) = \frac{\pi_\theta(a_t|s_t)}{\pi_{\theta_{old}}(a_t|s_t)}$ indicates a probability ratio. This procedure results in a clipping of the probability ratio $R_t(\theta)$ at time $1-\epsilon$ or $1+\epsilon$, depending upon whether the advantage is positive or negative, which forms the clipped objective after multiplying the advantage approximator $\hat{A}_t$. In the end, $L_t^{CLIP}$ is calculated by taking the minimum of this clipped objective and the unclipped objective $R_t(\theta)\hat{A}_t$, thereby reducing the need for a substantial policy update compared to the unclipped version [17]. This loss function is known as the conservative policy iteration algorithm loss function [21].

Using gradient descent, the parameters of $\pi_\theta$ are updated according to the gradient $\nabla L^{CLIP}$ of the negative of the clipped objective function:

$$\theta = \theta - \eta_\theta \nabla L^{CLIP}(\theta) \tag{10}$$

Assume that $\eta_\theta$ is the learning rate for the actor model optimization.

## IV. Proposed DRL Based Fault Diagnosis

### A. Data Pre-processing

**Turbine Databases** Some rows with a "drop-it" title are available in the datasets available. As such data is considered outlier data, it should be deleted. Following that, the data must be normalized using a Min-Max scaler as follows:

$$\mathrm{D}_{Normalize} = \frac{\mathrm{D} - \mathrm{D}_{min}}{\mathrm{D}_{max} - \mathrm{D}_{min}}. \tag{11}$$

$D$ and $D_{Normalize}$ represent features before and after normalization, respectively, and $D_{min}$ and $D_{max}$ represent minimum and maximum features based on data before normalization.

**CICIDS2017 Database.** This dataset is also normalized using the same technique as the previous dataset. In addition, some values had inf values, which posed a problem in the normalization and classification procedure. These values have been dropped from the list.

### B. Agent Architecture Design

We have faced many challenges in dealing with the imbalanced classification problem. In industrial settings, it is essential to detect faulty data, and many researchers have attempted to reduce the number of false positives and false negatives. The loss functions were modified to emphasize the few data that address the imbalanced data problem. Consider the objective function in the actor model. As follows, we define a new surrogate objective function:

$$L^{NCPI}(\theta) = \hat{E}_t \left[ \frac{\pi_\theta(a_t \mid s_t)}{\pi_{\theta_{old}}(a_t \mid s_t)} \hat{A}_t + \beta \log \left( \pi_\theta(a_t \mid s_t) \right) \right] \tag{12}$$

New Conservative Policy Iteration is indicated by the superscript N C P I. The term entropy into a new policy was added to the cost function compared to equation 6. The purpose of this work is to place a greater emphasis on new policies rather than old ones and to improve and accelerate the convergence of actor losses and total losses. In the results section, we will compare the results with and without this change.

The large class imbalance observed during the training of dense detectors overwhelms the cross-entropy loss, as demonstrated in [58]. The authors propose adapting the loss function to down-weight easy examples and adding a modulating factor to the cross-entropy loss. Therefore, the following is the definition of focal loss:

$$\mathrm{FL}(p_\mathrm{t}) = -\alpha_\mathrm{t} (1 - p_\mathrm{t})^\gamma \log(p_\mathrm{t}) \tag{13}$$

$p \in [0,1]$ is the estimated probability according to the model. Specifically, $\alpha \in [0,1]$ is a weighting factor, and $\gamma \in [0,1]$ is a tunable focusing parameter. A modulating factor reduces the loss contribution from easy examples and extends the range of examples subject to low loss contributions.

For the above reasons, we replace the entropy loss with a focal loss in equation 8. Results show a significant improvement in performance and a rapid convergence in total loss. This change will also be revealed in the results section. Below is a definition of the new total loss:

$$L_t^{NCLIP+V+S}(\theta) = \widehat{\mathbb{E}}_t \left[ L_t^{NCLIP}(\theta) - c_1 L_t^V(\theta) + c_2 FL\left[\pi_\theta\right](s_t) \right] \tag{14}$$

In Algorithm 1 and Fig. 2, the learning phase is described in detail, and a simulated environment model with new cost functions is presented. The evaluation part is described in Algorithm 2.

## V. Experimental Verification and Analysis

### A. Data Description

**Turbine Database 1.** This study used data from a real working steam turbine. The turbine's data was collected over 124 days at one minute per day sampling rate. The total number of data collected was 207,361. After preprocessing and removing outlier data from the non-dominant operating mode, 190,635 data could be used. A total of 121,279 data points are included in the normal category, and 69,356 data points are included in the fault category. There are 31 features included in this program, such as condenser pressure, pressure on inlet valves, steam flow value, active and reactive power, and others. The problem was caused by a leak in one of the pressure valves, causing the entire system to fail. This study examines the system's behavior in its dominant operating mode, the high-pressure mode, and the data are sufficiently comprehensive to understand the system's behavior.

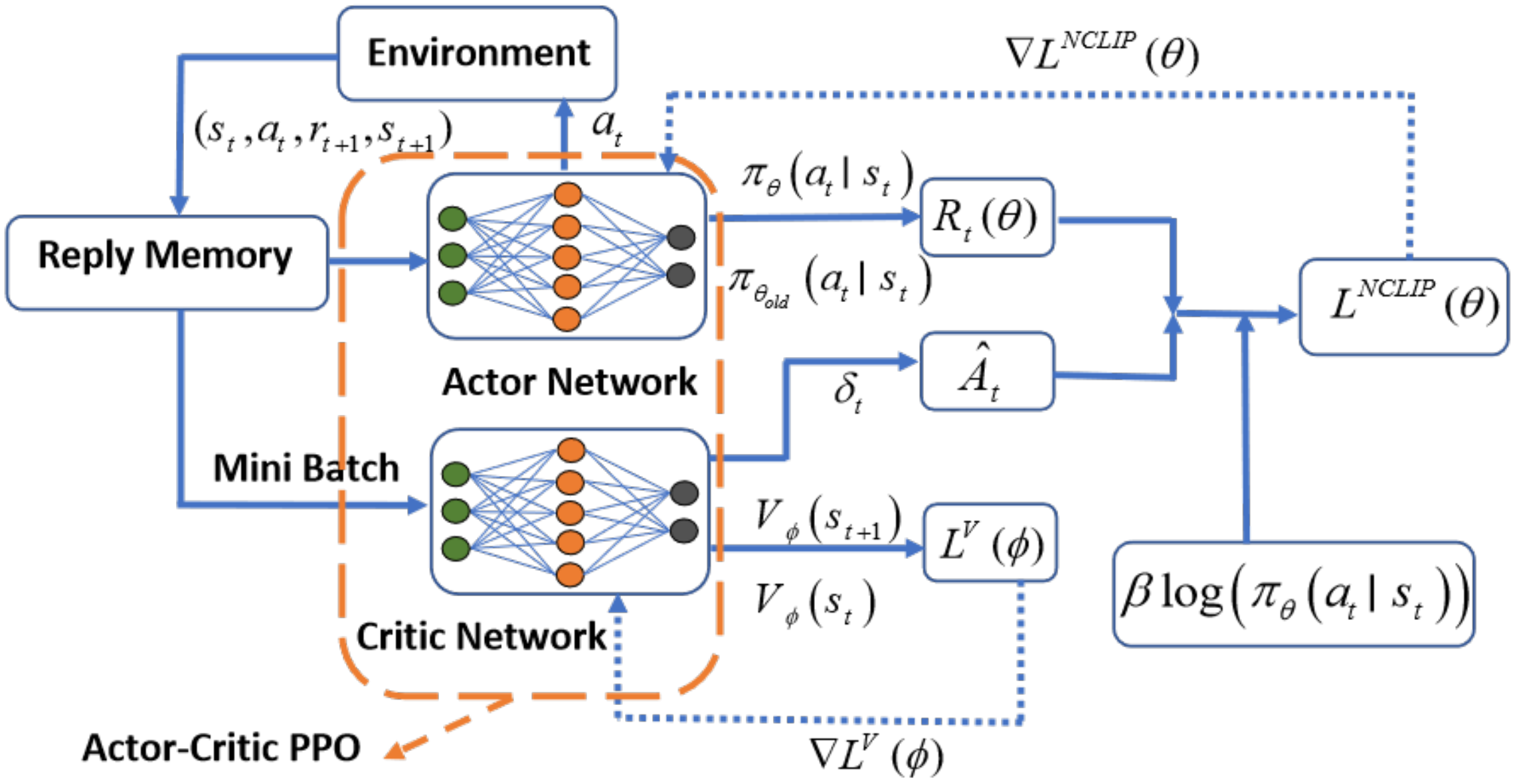

Fig. 2: Process of our actor-critic proximal policy optimization.

**Algorithm 1** Proposed Proximal Policy Optimization

**Require:**
States $\mathcal{S} = \{s_1, \ldots, s_n\}$
Actions $\mathcal{A} = \{a_1, \ldots, a_n\}$, $\quad A : \mathcal{S} \Rightarrow \mathcal{A}$
Reward function $R : \mathcal{S} \times \mathcal{A} \rightarrow \mathbb{R}$
Probabilistic transition function $P : \mathcal{S} \times \mathcal{A} \rightarrow \mathcal{S}$
Initialize ER buffer $B$ with experience replay memory $M$
**procedure** PROXIMAL POLICY OPTIMIZATION
  **for** each step of an episode **do**
    Run the actor model $\pi_\theta(s_t, a_t)$
    Store $(s_t, a_t, r_t, s_{t+1})$ in $B$
  **if** Learning is not finished **then**
    $\pi_{\theta_{\text{old}}} \leftarrow \pi_\theta$
    Random sample $(s_i, a_i, r_i, s_{i+1})$ from $M$
    Compute $\widehat{A_t}$ (using the critic model)
    Get the value $V_\phi(s_t)$, $\hat{V}_\phi^{\text{target}}(s_t)$
    Compute $L_t^V(\phi)$
    Compute $L_t^{NCLIP}(\theta)$ (using the actor model)
    Update critic model $V_\phi$ using $\nabla L^V(\phi)$
    Update actor model $\pi_\theta$ using $\nabla L^{NCLIP}(\theta)$
    Compute $FL[\pi_\theta](s_t)$
    Update policy $\pi$
  **if** Learning is finished **then**
    Store policy $\pi_\theta(a_t \mid s_t)$
    Evaluate $\pi$ on the test database

**Algorithm 2** Environment Evaluation

**Require:**
Labels $\mathcal{L} = \{l_1, \ldots, l_n\}$
Batch Samples $\mathcal{X} = \{(s_1, l_1), (s_2, l_2), \ldots, (s_T, l_B)\}$
**procedure** EVALUATION
  Our new policy $\pi(S) = \mathcal{A}$
  **for** sampled $\{s_k\}_{k=1}^B$ **do**
    **if** $a_t = l_t$ **then**
      $r_t = positive$
    **if** $a_t \neq l_t$ **then**
      $r_t = negative$
    Receive next state $s_{t+1}$
    **if** $t == B$ **then**
      Store variables and go to another batch.

**Turbine Database 2.** Another set of data with a different turbine is collected. The data was collected over 16 days at one minute per day sampling rate. Following preprocessing and removing outliers from the non-dominant operating mode, 240763 data in 8 classes could be used. The normal category contains 188,974 data points, while the fault category named ”rpm low” contains 5456 data points. There has been a decrease in turbine frequency, which has resulted in a decrease in turbine rotation speed, which is the cause of this fault. Another class has been labeled ”unknown,” and we have six different unknown labels. It was unclear to the expert man what the cause of the fault was. We have 32841, 6590, 5865, 593, and 285 records in each unknown class. This document, 65 features are included, such as compressor discharge pressure, active and reactive power of the generator, exhaust temperature, interstage fuel gas pressure, and others. This study examines the system’s behavior in its dominant operating mode, the L30 mode. In this operating mode, known as the temperature working mode, the turbine produces power to the maximum capacity allowed by the system. The absolute limit of the allowed value is the inlet temperature of the turbine (after combustion), which should be at most, a specific value to prevent overheating of the blades. The fuel valve controls the inlet turbine’s end temperature. This turbine has other working modes (start-up mode, coast-down mode, acceleration mode), but it works almost in this mode.

**CICIDS2017 Database.** The third dataset we used was the CICIDS2017 dataset [59], which represents external network data. This dataset contains the most up-to-date attack patterns in terms of network security. In order to speed up the evaluation process, we used one document from the database. As shown in Table I, the dataset is summarized.

TABLE I: Dataset summary for CICDS-2017.

| Filename | Classes | Samples |
|---|---|---|
| Wednesday working hours | Benign | 440031 |
| | DoS Hulk | 231073 |
| | DoS GoldenEye | 10293 |
| | DoS Slowloris | 5796 |
| | DoS Slowhttptest | 5499 |
| | Heartbleed | 11 |

### B. Evaluation Configuration

In order to implement the Actor–Critic PPO algorithm, Python 3.8 and PyTorch 1.11 are used. Actor and critic models are constructed using multilayer perceptrons of four hidden layers each and two separate output layers. For turbine database 1 and the CICIDS2017 database, each hidden layer consists of 32 neurons. For turbine database 2, we also use 48 neurons. The first output layer of the actor model results in several values that sum to one, for instance, eight actions for turbine database 2. As a result of the second output layer for the critic model, a single value can be obtained as an evaluation of the action selected by the actor model. Table II lists the other hyper-parameters of the Actor–Critic PPO algorithm.

The $\beta$ value in equation 12 is equal to 0.01. Equation 13 uses the $\alpha$ value of 0.25 and the $\gamma$ value of 2. Equation 14 uses 0.5 and 0.2 for the coefficients $c_1$ and $c_2$.

TABLE II: PPO training hyperparameters.

| The hyperparameter | Value |
|---|---|
| Optimization algorithm | Adam |
| GAE parameter | 0.95 |
| Clipping parameter | 0.2 |
| Learning rate (actor, critic) | 0.001 |
| Batch size | 256 |
| Discount factor | 0.99 |
| The number of steps | 256 |
| epoch | 100 |

### C. Results and Analysis

Based on the data descriptions in this section, the algorithm's productivity is evaluated on the above data. According to these results, the agent has learned a series of recognition strategies well. The agent can realize action learning quickly, mainly when performing diagnostic tasks related to turbine faults.

We considered three models to evaluate the PPO algorithm on our benchmarks, besides evaluating the changes in cost functions on the result. Firstly, we named traditional PPO "**Model 1**". As a second step, we changed entropy in equation 8 to focal loss, and we named our new model "**Model 2**". Lastly, after changing the cost function in the actor to equation 12 in addition to the use of focal loss and reach to equation 14, we named the final model "**Model 3**".

Fig. 3 illustrates that all evaluation criteria increased by approximately 3% with Model 2 compared to Model 1. In particular, Model 2 has an accuracy of 93.86%, the precision of 93.70%, recall of 92.94%, and f1 score of 93.32%. the values of evaluation criteria for Model 3 have increased to 94.84%, 94.88%, 94.84%, and 94.86%, respectively.

The change in the cost functions led to an increase in performance for the second dataset, as seen in Fig. 4. According to the evaluation criteria, Model 1 does not perform well. As a result of using Model 2, the accuracy, precision, recall, and F1 score are 98%, 78%, 88%, and 81%, respectively. Model 3's criteria values are 98%, 90%, 91%, and 90%, with an increase of 3%-12% in precision, recall, and F1 score over Model 2.

According to Fig. 5, traditional PPO performs at 85%, 93%, 85%, and 88% in criteria values based on the CICIDS2017 database. Model 2 increased all criteria values to 96%. As a final result in Model 3, all the criteria values were increased by about 3% rather than in Model 2 to 99%.

There is a significant increase in performance between Model 2 and Model 1, as can be seen in almost all figures. The performance of Model 3 has also been improved in comparison to Model 2 in terms of evaluation criteria. Due to the reasons we mentioned at the beginning of the article, fault detection is critical in industrial systems. The lack of labeled data in the first and second datasets, which relate to a real turbine, and the third dataset, which relates to an attack on a real system, make fault detection challenging. By enhancing performance in all benchmarks studied, the third model has solved the challenge of an imbalanced problem.

Table III compares our results on Turbine Database 1 with our previous work on this data [60]. Values of evaluation criteria were collected using weighted averages. The result we obtained with modified proximal policy is higher than that obtained with another RL framework named DDQN with Update Policy. In DDQN with Update Policy, we use double deep Q networks in a classification Markov decision process, and we periodically update the policy based on the data we receive. In this mode, we must update the initial model 124 times (because our data was gathered over 124 days), and each time, the model is run in 1000 iterations. Using our new method based on PPO, we achieve better performance in less time, and we only train our model on 100 epochs. As a result, with our new algorithm, we have maintained the performance while reducing computational time and cost.

To verify the validity of the proposed method and see the results and compare it with previous works, we considered one of the reliable datasets in the field of cyberattacks. Now we can see the advantages of the cost function changes compared to the earlier methods. It can be seen from Table III that our method performs better than contrastive learning-based methods [61], [62] and is comparable to the LSTM approach used in [63].

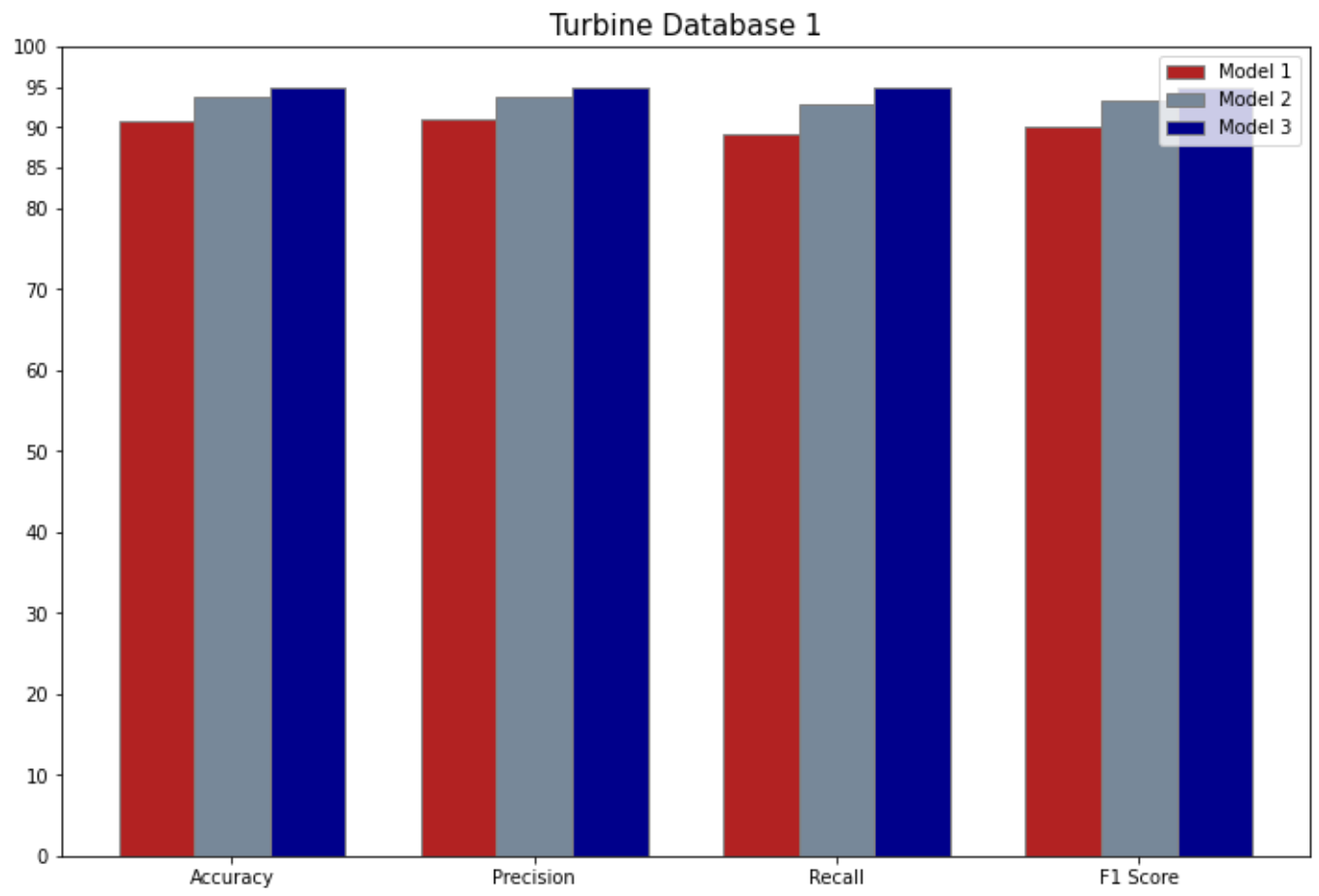

Fig. 3: Performance comparison of Models on Turbine Database 1.

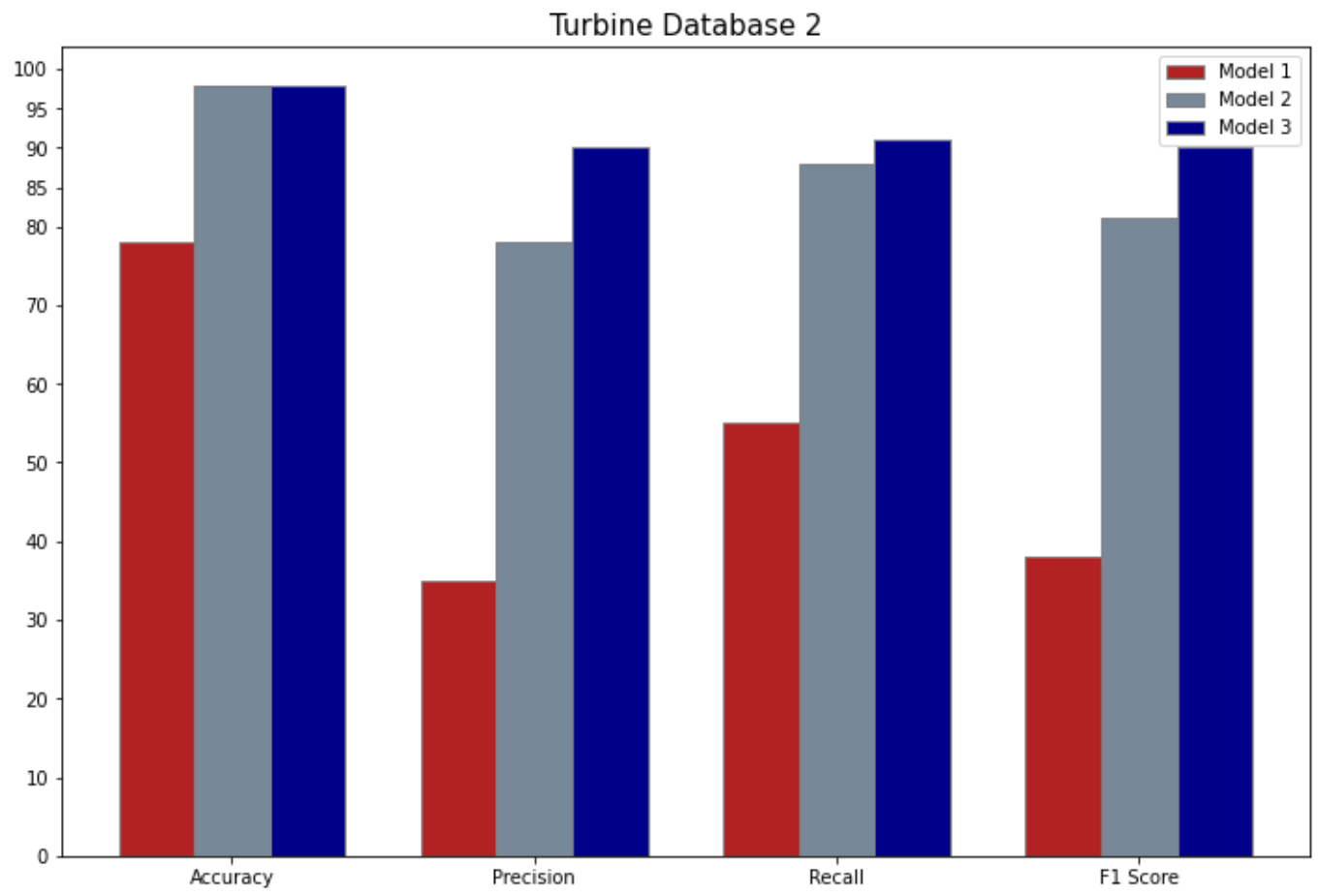

Fig. 4: Performance comparison of Models on Turbine Database 2.

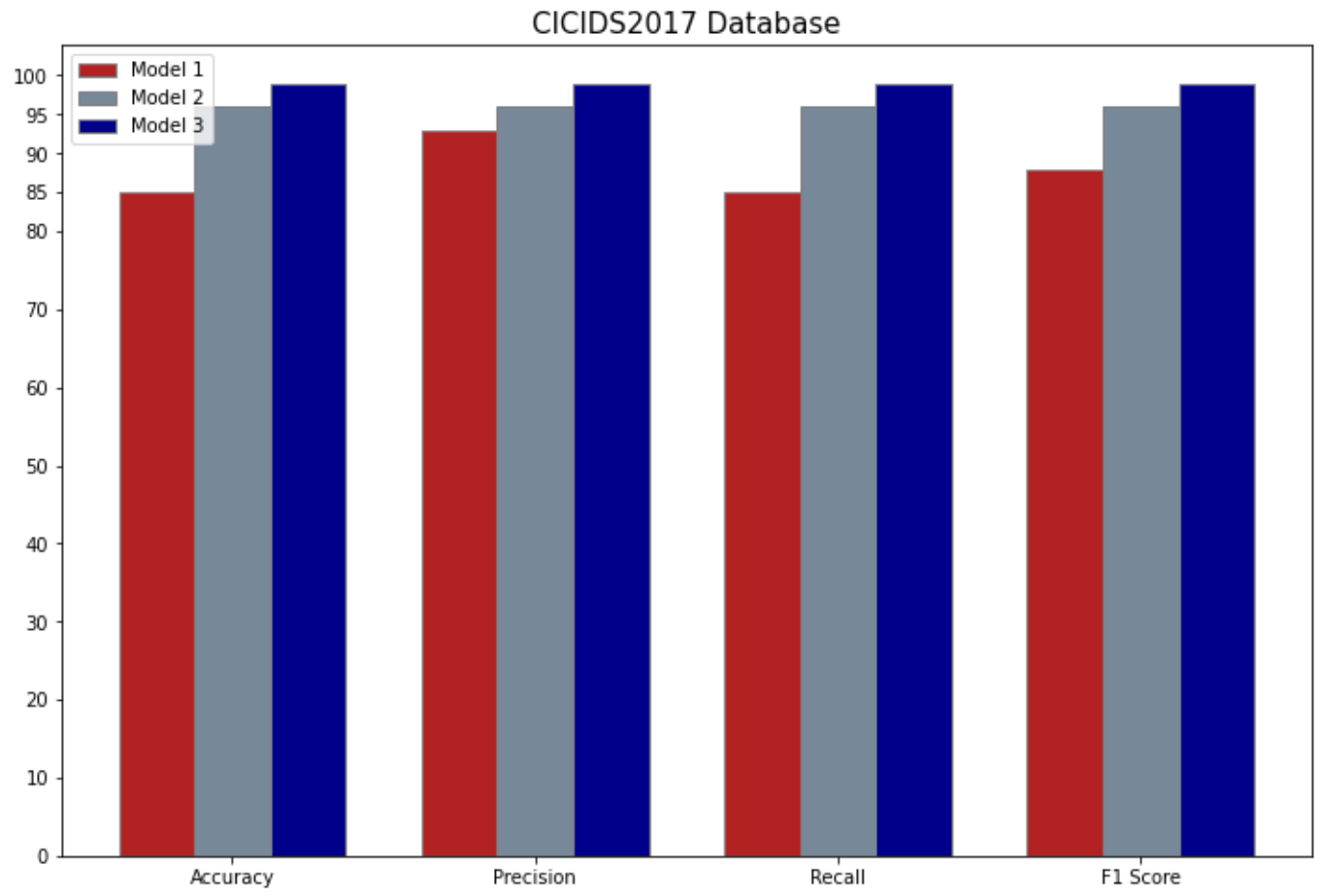

Fig. 5: Performance comparison of Models on CICIDS2017 Database.

TABLE III: Comparison of our PPO policy with prior work.

| Data | method | Accuracy | Precision | Recall | F1 Score |
|---|---|---|---|---|---|
| Turbine Database 1 | DDQN without Update Policy [60] | 0.91 | 0.91 | 0.91 | 0.91 |
| | DDQN with Update Policy [60] | 0.94 | 0.94 | 0.94 | 0.94 |
| | Our Proximal Policy Optimization | **0.95** | **0.95** | **0.95** | **0.95** |
| CIC IDS2017 | BoTNet [61] | 0.97 | 0.95 | 0.96 | 0.96 |
| | Contrastive learning [62] | 0.98 | 0.98 | 0.98 | 0.98 |
| | Refined LSTM [63] | **0.99** | **0.99** | **0.99** | **0.99** |
| | Our Proximal Policy Optimization | **0.99** | **0.99** | **0.99** | **0.99** |

## VI. Conclusions

In this paper, we present a method for fault detection based on neural networks and reinforcement learning. Our approach is based on a famous policy based on actor and critic networks, known as Proximal Policy Optimization. A disadvantage of the traditional policy is its inability to detect fault classes due to the lack of fault data, which was addressed by changing the cost function in actor loss and total loss. We observe a significant improvement in all evaluation criteria, as shown in the figures, and results are comparable to previous works in the first and third benchmarks. In addition, despite the second benchmark having eight imbalance classes, the new policy classifies it well, while the old policy does not. In addition, our new policy is more reasonable in terms of time and computational cost than our previous policy with the Double Deep Q Network in the first benchmark.

## References

[1] X. Cheng and J. Yu, "Retinanet with difference channel attention and adaptively spatial feature fusion for steel surface defect detection," *IEEE Transactions on Instrumentation and Measurement*, vol. 70, pp. 1–11, 2020.

[2] M. Miao and J. Yu, "A deep domain adaptative network for remaining useful life prediction of machines under different working conditions and fault modes," *IEEE Transactions on Instrumentation and Measurement*, vol. 70, pp. 1–14, 2021.

[3] Z. Gao, C. Cecati, and S. X. Ding, "A survey of fault diagnosis and fault-tolerant techniques—part i: Fault diagnosis with model-based and signal-based approaches," *IEEE transactions on industrial electronics*, vol. 62, no. 6, pp. 3757–3767, 2015.

[4] Y. Wang, Y. Si, B. Huang, and Z. Lou, "Survey on the theoretical research and engineering applications of multivariate statistics process monitoring algorithms: 2008–2017," *The Canadian Journal of Chemical Engineering*, vol. 96, no. 10, pp. 2073–2085, 2018.

[5] W. Yang, Z. Lang, and W. Tian, "Condition monitoring and damage location of wind turbine blades by frequency response transmissibility analysis," *IEEE Transactions on Industrial Electronics*, vol. 62, no. 10, pp. 6558–6564, 2015.

[6] Z. Elouedi, K. Mellouli, and P. Smets, "Assessing sensor reliability for multisensor data fusion within the transferable belief model," *IEEE Transactions on Systems, Man, and Cybernetics, Part B (Cybernetics)*, vol. 34, no. 1, pp. 782–787, 2004.

[7] H. Guo, W. Shi, and Y. Deng, "Evaluating sensor reliability in classification problems based on evidence theory," *IEEE Transactions on Systems, Man, and Cybernetics, Part B (Cybernetics)*, vol. 36, no. 5, pp. 970–981, 2006.

[8] Z. Gao and S. Sheng, "Real-time monitoring, prognosis, and resilient control for wind turbine systems," pp. 1–4, 2018.

[9] B. Yang, R. Liu, and X. Chen, "Fault diagnosis for a wind turbine generator bearing via sparse representation and shift-invariant k-svd," *IEEE Transactions on Industrial Informatics*, vol. 13, no. 3, pp. 1321–1331, 2017.

[10] E. Alizadeh, N. Meskin, and K. Khorasani, "A dendritic cell immune system inspired scheme for sensor fault detection and isolation of wind turbines," *IEEE Transactions on Industrial Informatics*, vol. 14, no. 2, pp. 545–555, 2017.

[11] H. Shao, Z. Gao, X. Liu, and K. Busawon, "Parameter-varying modelling and fault reconstruction for wind turbine systems," *Renewable Energy*, vol. 116, pp. 145–152, 2018.

[12] L. Wang, Z. Zhang, H. Long, J. Xu, and R. Liu, "Wind turbine gearbox failure identification with deep neural networks," *IEEE Transactions on Industrial Informatics*, vol. 13, no. 3, pp. 1360–1368, 2016.

[13] P. B. Dao, W. J. Staszewski, T. Barszcz, and T. Uhl, "Condition monitoring and fault detection in wind turbines based on cointegration analysis of scada data," *Renewable Energy*, vol. 116, pp. 107–122, 2018.

[14] X. Liu, Z. Gao, and M. Z. Chen, "Takagi–sugeno fuzzy model based fault estimation and signal compensation with application to wind turbines," *IEEE Transactions on Industrial Electronics*, vol. 64, no. 7, pp. 5678–5689, 2017.

[15] G. Wang and J. Jiao, "A kernel least squares based approach for nonlinear quality-related fault detection," *IEEE Transactions on Industrial Electronics*, vol. 64, no. 4, pp. 3195–3204, 2016.

[16] G. Wang, J. Jiao, and S. Yin, "A kernel direct decomposition-based monitoring approach for nonlinear quality-related fault detection," *IEEE Transactions on Industrial Informatics*, vol. 13, no. 4, pp. 1565–1574, 2016.

[17] R. K. Pandit, D. Infield, and A. Kolios, "Comparison of advanced non-parametric models for wind turbine power curves," *IET Renewable Power Generation*, vol. 13, no. 9, pp. 1503–1510, 2019.

[18] L. Yang and Z. Zhang, "Wind turbine gearbox failure detection based on scada data: A deep learning-based approach," *IEEE Transactions on Instrumentation and Measurement*, vol. 70, pp. 1–11, 2020.

[19] N. Huang, Q. Chen, G. Cai, D. Xu, L. Zhang, and W. Zhao, "Fault diagnosis of bearing in wind turbine gearbox under actual operating conditions driven by limited data with noise labels," *IEEE Transactions on Instrumentation and Measurement*, vol. 70, pp. 1–10, 2020.

[20] Y. Cui, P. Bangalore, and L. B. Tjernberg, "An anomaly detection approach based on machine learning and scada data for condition monitoring of wind turbines," in *2018 IEEE International Conference on Probabilistic Methods Applied to Power Systems (PMAPS)*, pp. 1–6. IEEE, 2018.

[21] Z. Wang, L. Wang, and C. Huang, "A fast abnormal data cleaning algorithm for performance evaluation of wind turbine," *IEEE Transactions on Instrumentation and Measurement*, vol. 70, pp. 1–12, 2020.

[22] M. Tan and Z. Zhang, "Wind turbine modeling with data-driven methods and radially uniform designs," *IEEE Transactions on Industrial Informatics*, vol. 12, no. 3, pp. 1261–1269, 2016.

[23] W. Sun, R. Zhao, R. Yan, S. Shao, and X. Chen, "Convolutional discriminative feature learning for induction motor fault diagnosis," *IEEE Transactions on Industrial Informatics*, vol. 13, no. 3, pp. 1350–1359, 2017.

[24] P. Bangalore and L. B. Tjernberg, "An approach for self evolving neural network based algorithm for fault prognosis in wind turbine," in *2013 IEEE Grenoble Conference*, pp. 1–6. IEEE, 2013.

[25] J. Liu, F. Qu, X. Hong, and H. Zhang, "A small-sample wind turbine fault detection method with synthetic fault data using generative adversarial nets," *IEEE Transactions on Industrial Informatics*, vol. 15, no. 7, pp. 3877–3888, 2018.

[26] J. Zeng, D. Lu, Y. Zhao, Z. Zhang, W. Qiao, and X. Gong, "Wind turbine fault detection and isolation using support vector machine and a residual-based method," in *2013 American control conference*, pp. 3661–3666. IEEE, 2013.

[27] B. Tang, T. Song, F. Li, and L. Deng, "Fault diagnosis for a wind turbine transmission system based on manifold learning and shannon wavelet support vector machine," *Renewable Energy*, vol. 62, pp. 1–9, 2014.

[28] M. Schlechtingen, I. F. Santos, and S. Achiche, "Wind turbine condition monitoring based on scada data using normal behavior models. part 1: System description," *Applied Soft Computing*, vol. 13, no. 1, pp. 259–270, 2013.

[29] Y. Lei, F. Jia, J. Lin, S. Xing, and S. X. Ding, "An intelligent fault diagnosis method using unsupervised feature learning towards mechanical big data," *IEEE Transactions on Industrial Electronics*, vol. 63, no. 5, pp. 3137–3147, 2016.

[30] W. Zhang, C. Li, G. Peng, Y. Chen, and Z. Zhang, "A deep convolutional neural network with new training methods for bearing fault diagnosis under noisy environment and different working load," *Mechanical Systems and Signal Processing*, vol. 100, pp. 439–453, 2018.

[31] G. Jiang, P. Xie, H. He, and J. Yan, "Wind turbine fault detection using a denoising autoencoder with temporal information," *IEEE/Asme transactions on mechatronics*, vol. 23, no. 1, pp. 89–100, 2017.

[32] X. Tao, D. Zhang, Z. Wang, X. Liu, H. Zhang, and D. Xu, "Detection of power line insulator defects using aerial images analyzed with convolutional neural networks," *IEEE Transactions on Systems, Man, and Cybernetics: Systems*, vol. 50, no. 4, pp. 1486–1498, 2018.

[33] L. Guo, Y. Lei, S. Xing, T. Yan, and N. Li, "Deep convolutional transfer learning network: A new method for intelligent fault diagnosis of machines with unlabeled data," *IEEE Transactions on Industrial Electronics*, vol. 66, no. 9, pp. 7316–7325, 2018.

[34] S. Shukla and B. Singh, "Reduced current sensor based solar pv fed motion sensorless induction motor drive for water pumping," *IEEE Transactions on Industrial Informatics*, vol. 15, no. 7, pp. 3973–3986, 2018.

[35] P. Cao, S. Zhang, and J. Tang, "Preprocessing-free gear fault diagnosis using small datasets with deep convolutional neural network-based transfer learning," *Ieee Access*, vol. 6, pp. 26 241–26 253, 2018.

[36] H. He and E. A. Garcia, "Learning from imbalanced data," *IEEE Transactions on knowledge and data engineering*, vol. 21, no. 9, pp. 1263–1284, 2009.

[37] H. He and E. A. Garcia, "Robust neural network fault estimation approach for nonlinear dynamic systems with applications to wind turbine systems," *EEE Transactions on Industrial Informatics*, vol. 15, no. 12, pp. 6302–6312, 2019.

[38] W. Siriseriwan and K. Sinapiromsaran, "Adaptive neighbor synthetic minority oversampling technique under 1nn outcast handling." *Songklanakarin Journal of Science & Technology*, vol. 39, no. 5, 2017.

[39] Z.-H. Zhou and X.-Y. Liu, "Training cost-sensitive neural networks with methods addressing the class imbalance problem," *IEEE Transactions on knowledge and data engineering*, vol. 18, no. 1, pp. 63–77, 2005.

[40] W. W. Ng, J. Zhang, C. S. Lai, W. Pedrycz, L. L. Lai, and X. Wang, "Cost-sensitive weighting and imbalance-reversed bagging for streaming imbalanced and concept drifting in electricity pricing classification," *IEEE Transactions on Industrial Informatics*, vol. 15, no. 3, pp. 1588–1597, 2018.

[41] S. A. Fayaz, S. Jahangeer Sidiq, M. Zaman, and M. A. Butt, "Machine learning: An introduction to reinforcement learning," *Machine Learning and Data Science: Fundamentals and Applications*, pp. 1–22, 2022.

[42] V. Mnih, K. Kavukcuoglu, D. Silver, A. Graves, I. Antonoglou, D. Wierstra, and M. Riedmiller, "Playing atari with deep reinforcement learning. arxiv [preprint] 2013," *arXiv preprint arXiv:1312.5602*, 2021.

[43] M. Fu, A. Agrawal, A. A. Irissappane, J. Zhang, L. Huang, and H. Qu, "Deep reinforcement learning framework for category-based item recommendation," *IEEE Transactions on Cybernetics*, 2021.

[44] N. Rudin, H. Kolvenbach, V. Tsounis, and M. Hutter, "Cat-like jumping and landing of legged robots in low gravity using deep reinforcement learning," *IEEE Transactions on Robotics*, vol. 38, no. 1, pp. 317–328, 2021.

[45] M. A. Wiering, H. Van Hasselt, A.-D. Pietersma, and L. Schomaker, "Reinforcement learning algorithms for solving classification problems," in *2011 IEEE Symposium on Adaptive Dynamic Programming and Reinforcement Learning (ADPRL)*, pp. 91–96. IEEE, 2011.

[46] Y. Ding, L. Ma, J. Ma, M. Suo, L. Tao, Y. Cheng, and C. Lu, "Intelligent fault diagnosis for rotating machinery using deep q-network based health state classification: A deep reinforcement learning approach," *Advanced Engineering Informatics*, vol. 42, p. 100977, 2019.

[47] J. Huang, Q. Chang, and J. Arinez, "Deep reinforcement learning based preventive maintenance policy for serial production lines," *Expert Systems with Applications*, vol. 160, DOI https://doi.org/10.1016/j.eswa.2020.113701, p. 113701, 2020. [Online]. Available: https://www.sciencedirect.com/science/article/pii/S095741742030525X

[48] H. Wang, J. Xu, C. Sun, R. Yan, and X. Chen, "Intelligent fault diagnosis for planetary gearbox using time-frequency representation and deep reinforcement learning," *IEEE/ASME Transactions on Mechatronics*, vol. 27, no. 2, pp. 985–998, 2021.

[49] T. V. Phan, T. G. Nguyen, N.-N. Dao, T. T. Huong, N. H. Thanh, and T. Bauschert, "Deepguard: Efficient anomaly detection in sdn with fine-grained traffic flow monitoring," *IEEE Transactions on Network and Service Management*, vol. 17, DOI 10.1109/TNSM.2020.3004415, no. 3, pp. 1349–1362, 2020.

[50] E. Bøhn, E. M. Coates, S. Moe, and T. A. Johansen, "Deep reinforcement learning attitude control of fixed-wing uavs using proximal policy optimization," in *2019 International Conference on Unmanned Aircraft Systems (ICUAS)*, DOI 10.1109/ICUAS.2019.8798254, pp. 523–533, 2019.

[51] W. Koch, R. Mancuso, R. West, and A. Bestavros, "Reinforcement learning for uav attitude control," *ACM Transactions on Cyber-Physical Systems*, vol. 3, pp. 1 – 21, 2018.

[52] J. Hwangbo, I. Sa, R. Siegwart, and M. Hutter, "Control of a quadrotor with reinforcement learning," *IEEE Robotics and Automation Letters*, vol. 2, DOI 10.1109/LRA.2017.2720851, no. 4, pp. 2096–2103, 2017.

[53] M. Zolotukhin, S. Kumar, and T. Hämäläinen, "Reinforcement learning for attack mitigation in sdn-enabled networks," in *2020 6th IEEE Conference on Network Softwarization (NetSoft)*, DOI 10.1109/NetSoft48620.2020.9165383, pp. 282–286, 2020.

[54] M. Chen, H. K. Lam, Q. Shi, and B. Xiao, "Reinforcement learning-based control of nonlinear systems using lyapunov stability concept and fuzzy reward scheme," *IEEE Transactions on Circuits and Systems II: Express Briefs*, vol. 67, DOI 10.1109/TCSII.2019.2947682, no. 10, pp. 2059–2063, 2020.

[55] H.-K. Lim, J.-B. Kim, J.-S. Heo, and Y.-H. Han, "Federated reinforcement learning for training control policies on multiple iot devices," *Sensors*, vol. 20, DOI 10.3390/s20051359, no. 5, 2020. [Online]. Available: https://www.mdpi.com/1424-8220/20/5/1359

[56] H. Han, H. Kim, and Y. Kim, "An efficient hyperparameter control method for a network intrusion detection system based on proximal policy optimization," *Symmetry*, vol. 14, DOI 10.3390/sym14010161, no. 1, 2022. [Online]. Available: https://www.mdpi.com/2073-8994/14/1/161

[57] F. Ye, X. Cheng, P. Wang, C.-Y. Chan, and J. Zhang, "Automated lane change strategy using proximal policy optimization-based deep reinforcement learning," in *2020 IEEE Intelligent Vehicles Symposium (IV)*, DOI 10.1109/IV47402.2020.9304668, pp. 1746–1752, 2020.

[58] T.-Y. Lin, P. Goyal, R. Girshick, K. He, and P. Dollár, "Focal loss for dense object detection," in *Proceedings of the IEEE international conference on computer vision*, pp. 2980–2988, 2017.

[59] I. Sharafaldin, A. H. Lashkari, and A. A. Ghorbani, "Toward generating a new intrusion detection dataset and intrusion traffic characterization." *ICISSp*, vol. 1, pp. 108–116, 2018.

[60] M. Modirrousta, M. A. Shoorehdeli, M. Yari, and A. Ghahremani, "Dqlap: Deep q-learning recommender algorithm with update policy for a real steam turbine system," *arXiv preprint arXiv:2210.06399*, 2022.

[61] Z. Wang, Z. Li, J. Wang, and D. Li, "Network intrusion detection model based on improved byol self-supervised learning," *Security and Communication Networks*, vol. 2021, 2021.

[62] S. Lotfi, M. Modirrousta, S. Shashaani, S. Amini, and M. A. Shoorehdeli, "Network intrusion detection with limited labeled data," *arXiv preprint arXiv:2209.03147*, 2022.

[63] K. O. Adefemi Alimi, K. Ouahada, A. M. Abu-Mahfouz, S. Rimer, and O. A. Alimi, "Refined lstm based intrusion detection for denial-of-service attack in internet of things," *Journal of Sensor and Actuator Networks*, vol. 11, no. 3, p. 32, 2022.

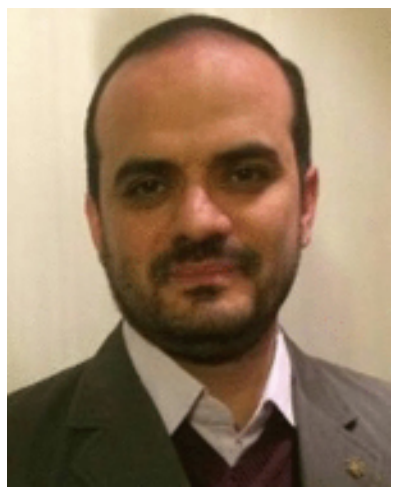

**Mahdi Aliyari Shoorehdeli** received his B.Sc. degree in electronics engineering from KNTU in 2001. He then pursued his studies in control engineering and, therefore, obtained his M.Eng. and Ph.D. degree from the same university in 2003 and 2008, respectively. He is currently appointed to the Department of Mechatronics Engineering of KNTU as an Assistant Professor. Dr. Aliyari is the author of more than 200 papers in international journals and conferences. His research interests include fault detection and isolation and system identification and Machine Learning.

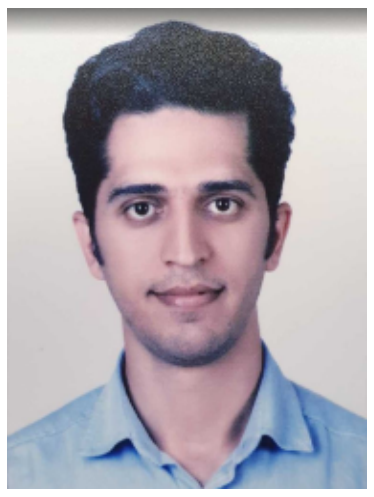

**Mostafa Yari** received his M.Eng. degree in mechatronics engineering from KNTU in 2013. His research interests include deep learning and fault detection and condition monitoring and image processing.

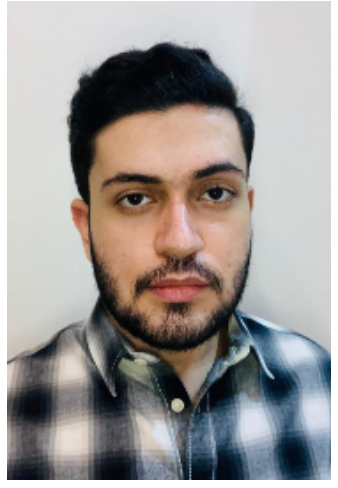

**Mohammad Hossein Modirrousta** received his B.Sc. degree in control engineering from KNTU in 2020. He then pursued his studies in control engineering and, therefore, obtained his M.Eng. degree from the same university in 2022, respectively. His research interests include deep learning and reinforcement learning and fault detection.

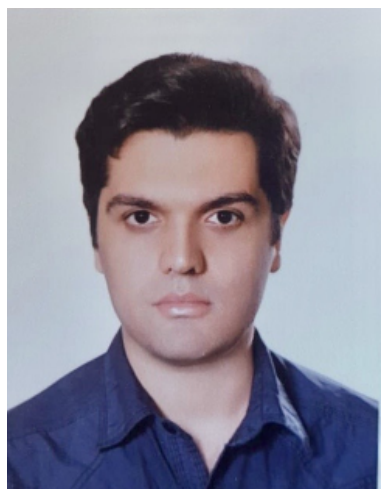

**Arash Ghahremani** received his M.Eng. degree in mechanical engineering (Energy Conversion) from KNTU in 2017. As a Ph.D. student, he is pursuing his studies in the same field and at the same university. His research interests include mechanics of fluids and solids, energy conversion and condition monitoring.